\documentclass[twocolumn,aps,showpacs,nofootinbib]{revtex4}
\usepackage{amsmath}
\usepackage{graphicx}
\usepackage{simon}

\begin{document}
\title{Polarization dependent photoionization cross-sections and radiative lifetimes of atomic states in Ba I}
\author{C.-H. Li}
\email{chli@mit.edu} \affiliation{Department of Physics,
University of California at Berkeley, Berkeley, California
94720-7300} \affiliation{Department of Physics, Massachusetts
Institute of Technology, Cambridge, Massachusetts 02139-4307}
\author{D. Budker}
\email{budker@berkeley.edu} \affiliation{Department of Physics,
University of California at Berkeley, Berkeley, California
94720-7300} \affiliation{Nuclear Science Division, Lawrence
Berkeley National Laboratory, Berkeley, California 94720}
\date{\today}
\begin{abstract}
The photoionization cross-sections of two even-parity excited
states, $5d6d~^3D_1$ and $6s7d~^3D_{2}$, of atomic Ba at the
ionization-laser wavelength of $556.6$~nm were measured. We found
that the total cross-section depends on the relative polarization
of the atoms and the ionization-laser light. With density-matrix
algebra, we show that, in general, there are at most three
parameters in the photoionization cross-section. Some of these
parameters are determined in this work. We also present the
measurement of the radiative lifetime of five even-parity excited
states of barium.
\end{abstract}
\pacs{32.10.-f,42.62.Fi}
\maketitle

\section{Introduction}

The photoionization cross-sections of atoms have been studied for
decades~\cite{Kel90}. 
With tunable lasers, it is possible to obtain high populations and
polarizations of selected excited states even if these states have
short lifetimes. We present here measurements of photoionization
cross-sections for excited states of Ba, which have been made
possible by this approach. The measurements of photoionization
cross-sections of atoms in excited states are valuable for testing
atomic theory, and are important for understanding of processes in
plasmas, including stellar atmospheres, lighting devices, etc. A
number of previous studies have observed that the photoionization
cross-section of polarized atoms depends on the polarization of
the light~\cite{Lub69, Fox71, Kog71}. In this work, we measured
the photoionization cross-sections and studied their polarization
dependence for the $5d6d~^3D_1$ and $6s7d~^3D_{2}$ states of
neutral barium at the ionization-laser wavelength of $556.6$~nm.
In addition to the applications mentioned above, our measurements
are also useful for the analysis of experiments with Ba searching
for violation of Bose-Einstein statistics (BEV) for
photons~\cite{Dem99, Eng00}.

\section{Experimental Method}
\label{Section:appratus}

\begin{figure}
\includegraphics[width=3.25 in]{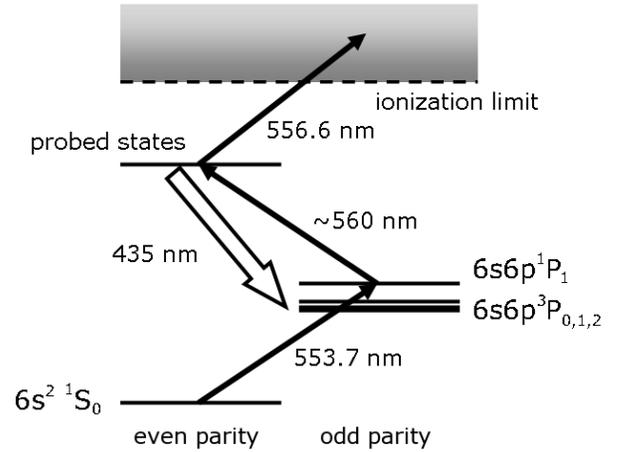}
\caption{The excitation-detection scheme. The probed states are
the $5d6d~^3D_1$ and $6s7d~^3D_{2}$ states. Solid arrows indicate
laser excitation; the hollow arrow indicates fluorescence. The
fluorescence is detected with a photomultiplier tube (PMT). The
ionization is detected by the induced charge on the electrodes,
which is converted to a voltage signal by a preamplifier.}
\label{fig:ionization}
\end{figure}

Two pulsed dye lasers are used to excite barium atoms in an atomic
beam to the even-parity states of interest via two successive E1
transitions. The barium atoms in the probed state can be ionized
by a third pulsed dye laser (Fig.~\ref{fig:ionization}).

\begin{figure}
\includegraphics[width=3.25 in]{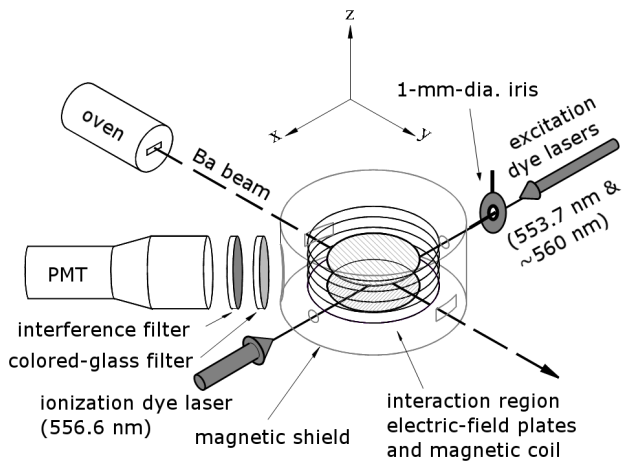}
\caption[Simplified schematic diagram of the apparatus.]
{Simplified schematic diagram of the apparatus.}
\label{fig:ionappartus}
\end{figure}

The apparatus used (Fig.~\ref{fig:ionappartus}) is largely the
same as in previous experiments~\cite{Roc99, Li04}. The barium
beam is produced with an effusive source with a multi-slit nozzle
that collimates the angular spread of the beam to $\sim 0.1$~rad
in both the horizontal and vertical directions. The oven,
heat-shielded with tantalum foil, is resistively heated to $\sim
700^{\circ}$C, corresponding to saturated barium pressure in the
oven of $\sim 0.1$~Torr and expected atomic-beam density in the
interaction region, $\sim 10$~cm away from the nozzle, of $\sim
10^{11}$~atoms/cm$^3$. However, the experimental estimate from the
fluorescence signal shows that the atomic density in the
interaction region is only $\sim 10^9$~atoms/cm$^3$ presumably
because of clogging in the nozzle. Residual-gas pressure in the
vacuum chamber is $\sim 2 \times 10^{-6}$~Torr.

The three tunable dye lasers used in this experiment (Quanta Ray
PDL-2, all with Fluorescein~548 dye) are pumped by two pulsed
frequency-doubled Nd-YAG lasers (Quanta Ray DCR-11 and Quantel
YAG580). The Quantel laser operates at a repetition rate of
$10$~Hz and slaves the Quanta Ray laser. The relative timing of
the two $\sim 7$-ns-long laser pulses can be controlled to within
$1$~ns. The output of the Quantel laser is split by a beam
splitter. One of the resultant light beams is used to pump the
(first) dye laser set on resonance with the $6s^2~^1S_0
\rightarrow 6s6p~^1P_1$ transition ($\lambda \sim 553.7$~nm). The
other light beam pumps the (third) dye laser, which is used to
ionize the barium atoms from the probed states. The wavelength of
the third dye laser is set to $556.6$~nm, which is relevant to the
BEV experiment. The (second) dye laser, set on resonance with the
transition from the $6s6p~^1P_1$ state to the probed state, is
pumped by the Quanta Ray Nd-YAG laser. The spectral width of each
of the dye-laser pulses is $\sim 20$~GHz. The spatial profile of
the dye-laser beam is approximately Gaussian, as measured with a
CCD camera. The Gaussian diameters of the laser beams in the
interaction region are adjusted to be $\sim 4$~mm. The relative
timing between the pulses of the first and second dye lasers
(excitation lasers) is set to maximize the population of the
probed state, at which point the pulses nearly coincide. The pulse
of the third dye laser (ionization laser) arrives in the
interaction region $\sim 30$~ns later, delayed by a spatial
distance. The excitation-laser beams are sent into the chamber in
the same direction, while the ionization-laser beam propagates in
the anti-parallel direction. The laser-beam paths are spatially
overlapped. An iris with a diameter of $1.01(2)$~mm is inserted
before the entrance of the excitation-laser beams (which is also
the exit of the ionization-laser beam), $\sim 50$~cm away from the
interaction region. The purpose of the iris is to control the
spatial distribution of the atoms in the probed states so that all
the atoms in the excited state are approximately uniformly
irradiated by the ionization-laser pulse.

The typical pulse energy of each of the excitation-laser beams is
$\sim 1.5$~mJ before they pass through the iris. The pulse energy
of the ionization-laser beam is $\sim 10$~mJ. A $1$-mm-thick
coated etalon is inserted in the ionization-laser-beam path before
the entrance of the beam into the chamber. We can adjust the pulse
energy of the ionization-laser beam by tilting the etalon. In
order not to change the laser-beam path significantly, we set the
etalon surface almost perpendicular to the laser-beam path (at an
angle $\le 5^{\circ}$), so the etalon parallel-shifts the beam by
less than $0.1$~mm. After the interaction region, the
ionization-laser beam passes through an exit window, with a
transmission rate of $89(2)\%$ and the $1.01$-mm iris, and is
split by a wedged piece of fused-silica glass. The energy of one
of the split beams is measured by a photodiode. To calibrate the
photodiode as an energy meter, we measure the energy of the
through beam (with energy $90(3)\%$ of that before the splitter) and 
the output voltage of the photodiode simultaneously, and fit them
to a linear function. The nonlinear deviation is found to be
$<2\%$ and is statistically negligible. We estimate the photon
flux density with the assumption that the intensity of the light
in the interaction region is homogenous and is proportional to
that of the light measured by the photodiode. The error due to
this assumption is $\sim 10\%$ in the photoionization
cross-section (see Section~\ref{Section:systematics}).

Fluorescence resulting from spontaneous decay to a lower-lying
state is detected at $45^{\circ}$ to both the atomic and
excitation-laser beams with a 2-in.-diameter PMT (EMI 9750B). The
gain of the PMT is $\sim 7\times 10^5$ (with an applied voltage of
$1.2$~kV), and the quantum efficiency at the wavelengths used is
$\sim 25\%$. Interference filters with $10$-nm bandwidth are used
to select decay channels of interest, and a colored-glass filter
is used to further reduce the scattered light from the lasers.

An electric field of $\sim 1$~kV/cm in the interaction region is
supplied by two plane-parallel electrodes. A detailed description
of the electrodes can be found in Ref.~\cite{Roc99}. The purpose
of the electric field is to separate the ions and the free
electrons, which are mutually attracted due to the induced
electric field of the space charge. The number of ions produced at
highest ionization light power is $\sim 2 \times 10^6$,
corresponding to a space-charge density of $\sim 2 \times
10^7~e$/cm$^2$, where $e$ is the charge of the electron, resulting
in an electric field of $\sim 30$~V/cm. We apply a field that is
larger than the space-charge field. On the other hand, the
electric field should be sufficiently low to avoid excessive
Stark-induced level mixing. An applied electric field of $1$~kV/cm
can cause $< 3\%$ of Stark-induced mixing for specific levels of
interests. Ions and free electrons are detected by the induced
charge on the electrodes, which is converted to a voltage signal
by a preamplifier (Tennelec TC174).

We use CAMAC modules connected through a general-purpose-interface
bus (GPIB) to a personal computer running LABVIEW software for
data acquisition. The fluorescence signal, the ion signal, and the
pulse energy of the ionization laser are recorded.

In this work, we study the dependence of the photoionization
cross-section on the relative polarizations between atoms and the
ionization laser. To produce different polarizations of the atoms
in the probed state, we vary the polarizations of the
excitation-laser beams with half-wave plates and purify the
polarizations with polarizers.

\begin{figure}
\includegraphics[width=3.25 in]{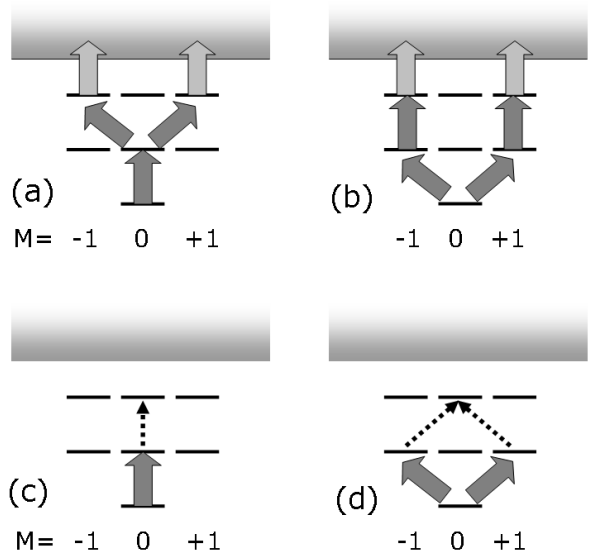}
\caption{The populated sublevels of a $J=1$ state according the
the polarizations of the excitation lasers. A solid arrow means an
allowed transition. A dashed arrow means a forbidden transition.
In the plots (a) and (b), it is shown that $M=\pm 1$ sublevels can
be excited when the polarization directions of the excitation
lasers are perpendicular to each other. In the plots (c) and (d),
it is shown that no sublevels can be excited when polarization
directions of the excitation lasers are parallel to each other
because the Clebsch-Gordan coefficient vanishes in this case. Case
(d) is physically equivalent to case (c); however, in this basis
the Clebsch-Gordan suppression of case (c) shows up as a
cancellation of coherent excitation paths via the $M=1$ and $M=-1$
sublevels.} \label{fig:J1plot}
\end{figure}

\begin{figure}
\includegraphics[width=3.25 in]{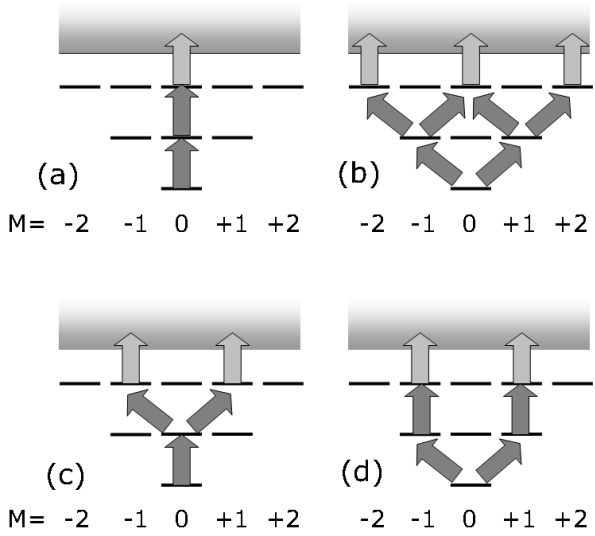}
\caption{The populated sublevels of a $J=2$ state according the
the polarizations of the excitation lasers. Plot (a) shows that
the $M=0$ sublevel can be excited if both lasers are polarized in
the $\hat{z}$ direction. Plot (b) shows that a coherent
superposition of $M= \pm 2$ and $M=0$ sublevels can be excited if
both lasers are polarized in the $\hat{y}$ direction. Plots (c)
and (d) show that a coherent superposition of $M=\pm 1$ sublevels
can be excited if one laser is polarized in the $\hat{z}$
direction and the other is polarized in the $\hat{y}$ direction.}
\label{fig:J2plot}
\end{figure}

We use the polarization of the linearly polarized ionization-laser
light to define the quantization axis $\hat{z}$. Because the
excitation and ionization laser beams propagate along the same
axis, defined as the $x$-axis, and light is transverse, the
polarization of the excitation-laser beams can only be in the
$\hat{y}$-$\hat{z}$ plane. For a $J=1$ probed state, only $M=\pm
1$ Zeeman sublevels can be coherently excited
(Fig.~\ref{fig:J1plot}). The $M=0$ sublevel cannot be excited
because the corresponding Clebsch-Gordan coefficient for a $J=1
\rightarrow J'=1$ transition is zero. For a $J=2$ probed state,
three different polarizations of the state can be obtained
(Fig.~\ref{fig:J2plot}). When both excitation lasers are polarized
along the $z$-axis, the $M=0$ sublevel is populated. When one of
the excitation lasers is polarized along the $z$-axis and the
other is polarized along the $y$-axis, $M=\pm 1$ sublevels are
coherently excited. When the polarizations of both lasers are
along the $y$-axis, $M= \pm 2$ and $M=0$ sublevels can be
coherently excited.

\section{Theory}

\subsection{Photoionization Cross-Section}

In this subsection, we show that there are at most three
parameters in the total photoionization cross-section using
density-matrix algebra (see, for example,
Ref.~\cite{Ale2005,Blum}). Assuming that the wavelength of the
ionization photons is much longer than the size of an atom, we
only consider electric-dipole transitions.

The density matrix describing an ensemble of atoms in a state with
total angular momentum $J$ can be expressed in the basis of its
$2J+1$ Zeeman components.
\begin{eqnarray}
\rho^{(i)}_a&=&\sum_{M M' = -J}^J \rho_{aMM'}^{(i)}
\ket{JM}\bra{JM'},\\
\rho_{a}&=&\frac{1}{N} \sum_{i=1}^N \rho^{(i)}_a,
\end{eqnarray}
where $\rho_{a~MM'}^{(i)}$ are the coefficients in the Zeeman
basis of the $i^{th}$ atom, and $\rho_{a}$ is the average density
matrix of an ensemble of $N$ atoms. As we will calculate the
photoionization cross-section which is normalized by the number of
particles, we likewise choose the normalization
\begin{equation}
Tr[\rho_a]=1.
\end{equation}

It is convenient to expand the density matrix in the basis of
irreducible tensors of rank $\kappa$ ($\kappa=0,1,\cdots,2J$),
\begin{equation}
\rho_a= \sum_{\kappa=0}^{2J}\sum_{q=-\kappa}^{\kappa}
\rho^{(\kappa)}_{a~q} T^{(\kappa)}_{q},
\end{equation}
where $T^{(\kappa)}$ are normalized polarization operators which
are irreducible tensors of rank $\kappa$ with $2\kappa+1$
components $T^{(\kappa)}_{q}$ ($q = - \kappa , -\kappa+1, \cdots ,
\kappa$), and $\rho^{(\kappa)}_{a~q}$ are coefficients, which are
related to $\rho_{aMM'}$ according to
\begin{equation}
\rho_{a~q}^{(\kappa)}=\sum_{MM'=-J}^J (-1)^{J-M'} \langle
J,M,J,-M'|\kappa,q\rangle \rho_{aMM'}, \label{Eqn:MMtoKQ}
\end{equation}
where $\langle J,M,J,-M'|\kappa,q\rangle$ are Clebsch-Gordan
coefficients.

Photons have total angular momentum $J=1$ in the electric-dipole
approximation. Therefore, the density matrix of the photons can be
decomposed into irreducible tensors of ranks $\kappa=0, 1$ and
$2$.
\begin{equation}
\rho_{p}= \sum_{\kappa=0}^{2}\sum_{q=-\kappa}^{\kappa}
\rho^{(\kappa)}_{p~q} T^{(\kappa)}_{q},
\end{equation}
where $\rho_p$ is normalized as $\rho_a$.

The photoionization process is related to the density matrices of
the ionizing photons and the probed state. Because total
photoionization cross-section is a scalar
(since we did not study the angular distribution of the ions), the
irreducible tensors of the density matrix of the photons should be
contracted with those of the atoms of the same ranks. The
photoionization cross-section can be expressed as:
\begin{equation}
\begin{split}
\sigma =  & \sqrt{3(2J+1)} \left(   \sigma_0
\rho^{(0)}_{p~0}\rho^{(0)}_{a~0} +  \sigma_1 \sum_{q=-1}^1 (-1)^q
\rho^{(1)}_{p~q} \rho^{(1)}_{a~-q} \right. \\
& \left. + \sigma_2 \sum_{q=-2}^2 (-1)^q \rho^{(2)}_{p~q}
\rho^{(2)}_{a~-q} \right),
\label{Eqn:cs3p}
\end{split}
\end{equation}
where $\sigma_{0,1,2}$ are coefficients determined by the atomic
wavefunctions of the initial and final (continuum) states. The
normalization factor $\sqrt{3(2J+1)}$ is chosen because according
to Eq.~(\ref{Eqn:MMtoKQ}),
\begin{eqnarray}
\rho_{a~0}^{(0)}&=&\frac{1}{\sqrt{2J+1}},\\
\rho_{p~0}^{(0)}&=&\frac{1}{\sqrt{3}}.
\end{eqnarray}
Therefore, we conclude that in general there are at most three
parameters in the photoionization cross-section. (There is only
one parameter for a $J=0$ state and there are two parameters for a
$J=1/2$ state.) For an unpolarized initial atomic state or/and an
unpolarized ionization light source,\footnote{Here the unpolarized
light source means that the diagonal elements of the density
matrix of the light are equal and the off-diagonal elements are
all zero. A directional light beam cannot be unpolarized in this
definition because of the lack of the polarization along its
propagation direction. Such light that can be ``unpolarized" in
the sense of the common definition through Stokes' parameters, in
fact, possesses alignment along the propagation direction.} the
photoionization cross-section is $\sigma_0$. If both of atoms and
light are polarized, the cross-section may be different depending
on their relative orientation ($\sigma_1$) and their relative
alignment ($\sigma_2$).

\subsection{Formulae for $\sigma_{0,1,2}$}

In this subsection, we derive general formulae for
$\sigma_{0,1,2}$ for states of arbitrary angular momenta. Consider
an ensemble of atoms prepared in a particular Zeeman sublevel of the
probed state $\ket{a J M}$, where $a$ represents all other quantum
numbers of the state, which are ionized by left-circularly polarized
photons. The density matrix of the photons is
\begin{equation}
\begin{array}{c}
\begin{array}{ccccc}
&~~~~~~ & M=+1 & ~M=0 & M=-1
\end{array}\\
\begin{array}{c}
M=+1\\
M=0\\
M=-1
\end{array}
\left( \begin{array}{ccc}
~~~~~1~~~~~ & ~~~~~0~~~~~ & ~~~~~0~~~~~ \\
0  & 0 & 0\\
0 & 0 & 0 \\
\end{array} \right).
\end{array}
\label{Eqn:mixingsecular}
\end{equation}
Using Eq.~(\ref{Eqn:MMtoKQ}), we can decompose it into irreducible
tensors with components $\rho_{p~q}^{(\kappa)}$:
\begin{equation}
\begin{array}{cc}
\kappa & (q=-\kappa,\cdots,\kappa)\\
0 & \left(\frac{1}{\sqrt{3}}\right)\\
1 & \left(0,\frac{1}{\sqrt{2}},0\right)\\
2 & \left(0,0,\frac{1}{\sqrt{6}},0,0\right).
\end{array}
\end{equation}
All elements of the density matrix of the probed state are zero
except $\rho_{aMM}=1$. Using Eq.~(\ref{Eqn:MMtoKQ}), we can
decompose it into irreducible tensors with components
$\rho_{a~q}^{(\kappa)}$:
\begin{equation}
\begin{array}{cc}
\kappa & (q=-\kappa,\cdots,\kappa)\\
0 & \left((-1)^{J-M} \threej(J,M)(J,-M)(0,0)
\right)\\
1 & \left(0,(-1)^{J-M} \sqrt{3} \threej(J,M)(J,-M)(1,0) ,0 \right)\\
2 & \left(0,0,(-1)^{J-M} \sqrt{5}\threej(J,M)(J,-M)(2,0) ,0,0
\right).
\end{array}
\end{equation}
According to Eq.~(\ref{Eqn:cs3p}) the photoionization
cross-section is
\begin{equation}
\begin{split}
\sigma=&\sqrt{3(2J+1)}\left(\sigma_0
\frac{(-1)^{J-M}}{\sqrt{3}} \threej(J,M)(J,-M)(0,0) \right.\\
&+\sigma_1(-1)^{J-M}\sqrt{\frac{3}{2}} \threej(J,M)(J,-M)(1,0)\\
&\left.+\sigma_2(-1)^{J-M}\sqrt{\frac{5}{6}}\threej(J,M)(J,-M)(2,0)
\right).
\end{split}\label{Eqn:sigma_exp1}
\end{equation}

It is convenient to introduce a function $Z$ defined as:
\begin{equation}
Z(l)=\sum_{M=-J}^{J} (-1)^{J+M}\threej(J,M)(J,-M)(l,0) \sigma.
\label{Eqn:Z}
\end{equation}
Using the identity of
\begin{equation}
\sum_{M=-J}^{J} \threej(J,M)(J,-M)(\kappa,0)
\threej(J,M)(J,-M)(l,0) = \frac{1}{2l+1}\delta_{\kappa l},
\end{equation}
it can be shown that
\begin{eqnarray}
Z(l=0)&=&\sqrt{2J+1}\sigma_0,\nonumber\\
Z(l=1)&=&\sqrt{\frac{2J+1}{2}}\sigma_1, \label{Eqn:Zsigma} \\
Z(l=2)&=&\sqrt{\frac{2J+1}{10}}\sigma_2.\nonumber
\end{eqnarray}

From Ref.~\cite{Sobelman}, the photoionization cross-section can
be expressed as
\begin{equation}
\sigma=\frac{4 \pi^2 m_e}{\hbar^2} \frac{k}{p} \sum_{n}
|\matelem{\psi_{n}}{\bf D^1_q}{\psi_a}|^2,\label{Eqn:cs eq}
\end{equation}
where $m_e$ is the mass of the electron, $k$ is the momentum of an
ionizing photon, $p$ is the momentum of an ionized electron,
$\psi_{a}$ is the atomic wavefunction of the probed state and
$\psi_{n}$ is the wavefunction of the coupled continuum state.

Using the relation
\begin{eqnarray}
&&|\matelem{n J_n M+1}{\rm \bf D^{(1)}_1 }{a J M}|^2 \nonumber \\
&=&-
\threej(J_n,-M-1)(1,1)(J,M)\threej(J,-M)(1,-1)(J_n,M+1)\nonumber\\
&& \cdot |\rme{n J_n}{{\rm \bf D}}{a J}|^2,
\end{eqnarray}
the photoionization cross-section in the example we are considering
can be written as
\begin{equation}
\sigma = -\sum_{n} A_n \cdot
\threej(J_n,-M-1)(1,1)(J,M)\threej(J,-M)(1,-1)(J_n,M+1),\label{Eqn:sigma_exp2}
\end{equation}
where $A_n=\frac{4 \pi^2 m_e k}{\hbar^2 p} \cdot |\rme{n J_n}{{\rm
\bf D}}{a J}|^2$. Using Eq.~(\ref{Eqn:sigma_exp2}) to calculate
$Z$ defined in Eq.~(\ref{Eqn:Z}),
\begin{eqnarray}
Z(l)&=&\sum_n A_n \cdot \sum_M (-1)^{J+M+1}\threej(J_n,-M-1)(1,1)(J,M)\nonumber\\
&&\cdot \threej(J,-M)(1,-1)(J_n,M+1)\threej(J,M)(J,-M)(1,0)\nonumber\\
&=&\sum_n A_n \cdot (-1)^{J-J_n}
\threej(1,1)(1,-1)(1,0)\sixj(1,1,l)(J,J,J_n),~~~
\label{Eqn:Zsigma2}
\end{eqnarray}
where, in the last line, we have used the identity
\begin{eqnarray}
&&\sum_{m_4 m_5 m_6}(-1)^{j_4+j_5+j_6-m_4-m_5-m_6}
\threej(j_1,m_1)(j_5,-m_5)(j_6,m_6)\nonumber\\
&&\cdot\threej(j_4,m_4)(j_2,m_2)(j_6,-m_6)\threej(j_4,-m_4)(j_5,m_5)(j_3,m_3)\nonumber\\
&=&\threej(j_1,m_1)(j_2,m_2)(j_3,m_3)\sixj(j_1,j_2,j_3)(j_4,j_5,j_6).
\end{eqnarray}

Comparing Eq.~(\ref{Eqn:Zsigma2}) with Eq.~(\ref{Eqn:Zsigma}), we
can derive the formulae for $\sigma_{0,1,2}$,
\begin{eqnarray}
\sigma_0&=&\sum_n A_n \cdot \frac{(-1)^{1-2J_n}}{3(2J+1)},\\
\sigma_1&=&\sum_n A_n \cdot
\frac{(-1)^{J-J_n}}{\sqrt{3(2J+1})}\sixj(1,1,1)(J,J,J_n),\\
\sigma_2&=&\sum_n A_n \cdot
\frac{(-1)^{J+J_n}}{\sqrt{3(2J+1})}\sixj(1,1,2)(J,J,J_n).
\end{eqnarray}
If the probed state is dominantly coupled to continuum states with
a specific total angular momentum $J_n=J_c$, the ratios between
photoionization cross-sections are
\begin{eqnarray}
\frac{\sigma_1}{\sigma_0}&=&(-1)^{J+J_c-1}\sqrt{3(2J+1)}\sixj(1,1,1)(J,J,J_c),\\
\frac{\sigma_2}{\sigma_0}&=&(-1)^{J-J_c-1}\sqrt{3(2J+1)}\sixj(1,1,2)(J,J,J_c).
\end{eqnarray}
In Table~\ref{table:coupling J}, we list the ratios between
$\sigma_{0,1,2}$ of a probed state with total angular momentum
$J=1,2$ if it is dominantly coupled to continuum states with total
angular momentum $J_{c}$.
\begin{table}[tbp]
\begin{center}
\begin{tabular}{cc|ccccc}
\hline \hline
$J$ & $J_{c}$ & ~~~~$\sigma_0$~~~~ &:& ~~~~$\sigma_1$~~~~ & :& ~~~~$\sigma_2$~~~~  \\
\hline
 &  0 & 1&: & -1&: & 1 \\
1 &  1 & 1 &:& -$\frac{1}{2}$&: & -$\frac{1}{2}$ \\
 &  2 & 1 &:& $\frac{1}{2}$ &:& $\frac{1}{10}$ \\
\hline
 &  1 & 1 &:& -$\sqrt{\frac{3}{4}}$ &:& $ \sqrt{\frac{7}{20}}$ \\
2 &  2 & 1 &:& -$\sqrt{\frac{1}{12}}$ &:& -$\sqrt{\frac{7}{20}}$ \\
 &  3 & 1 &:& $\sqrt{\frac{1}{3}}$ &:& $\sqrt{\frac{1}{35}}$ \\
\hline \hline
\end{tabular}
\caption{Ratios between photoionization cross-sections. The ratios
are calculated assuming that a probed state with total angular
momentum $J$ is dominantly coupled to continuum states with total
angular momentum $J_c$.} \label{table:coupling J}
\end{center}
\end{table}

\subsection{Ion-Signal Model}

In this subsection, we derive a formula describing the relation
between the ion signal and the photon fluence (the total photon
number per unit area) of the ionization-laser pulse, taking into
account the finite radiative lifetime of the probed state.

The change of the number of atoms ($N$) in the probed state is due
to the photoionization process and the spontaneous decay:
\begin{equation}
dN = - s(t) \sigma N dt - \frac{N}{\tau}dt, \label{Eqn:dN}
\end{equation}
where $s(t)$ is the temporal distribution of the photon number
intensity of the ionization-laser pulse, $\sigma$ is the
photoionization cross-section, and $\tau$ is the radiative
lifetime of the probed state. Assume that the ionization-laser
pulse comes into the interaction region at $t=0$ and $N(t=0)=N_0$.
Integrating Eq.~(\ref{Eqn:dN}), for $t>0$, we get
\begin{equation}
N(t) = N_0 e^{- \sigma \int_{0}^{t} s(t') dt'} e^{-t/\tau}.
\end{equation}
The number of ions detected after the pulse ($N_{\rm ion}(t
\rightarrow \infty)$) is
\begin{equation}
N_{\rm ion} = \int_{0}^{\infty} s(t) \sigma N(t) dt.
\end{equation}

We model the temporal distribution of the photon number intensity
of the ionization-laser pulse with a square function, i.e.
\begin{equation}
s(t)=\left\{ \begin{array}{ll} \frac{n}{\tau_l} & {\rm for~}
0<t<\tau_l,\\
0 & {\rm otherwise},\end{array} \right. \label{Eqn:temporalsquare}
\end{equation}
where $n$ is the photon fluence of the whole pulse and $\tau_l$ is
the duration of the pulse. We can then derive a formula relating
$N_{\rm ion}$ to $n$:
\begin{eqnarray}
N_{\rm ion}&=& N_0 \frac{n \sigma}{\tau_l} \int_{0}^{\tau_l} e^{-(\sigma n / \tau_l + 1/\tau) t} dt\nonumber\\
& = & N_0\frac{\sigma n}{\sigma n+ \tau_l / \tau} (1 - e^{
- \sigma n - \tau_l/ \tau}). \label{Eqn:ch4ion}
\end{eqnarray}
The error due to the assumption of Eq.~(\ref{Eqn:temporalsquare})
is estimated to be $\sim 4\%$ (see
Section~\ref{Section:systematics}).

\section{Results and Analysis}
\label{Section:results}

\begin{figure}
\includegraphics[width=3.25 in]{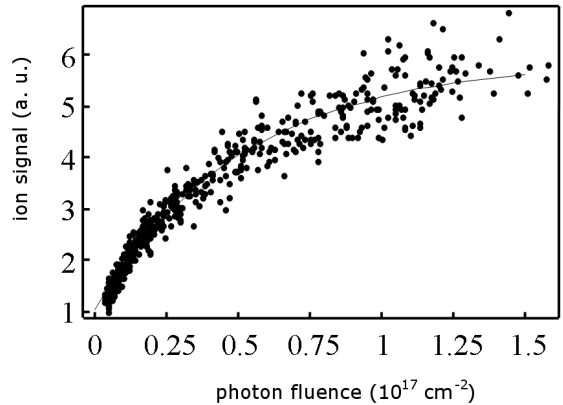}
\caption{The ion signal as a function the photon fluence of the
ionization laser pulse. The photon fluence is the time-integrated
photon-number intensity over a pulse duration. The atoms were in
the $5d6d~^3D_1$ state, excited via two E1 transitions, before
they were ionized. The data points are fit to Eq.~(\ref{Eqn:ion
fun}).} \label{fig:ionsignal}
\end{figure}

According to Eq.~(\ref{Eqn:ch4ion}), we determine the
photoionization cross-section, $\sigma$, by fitting the ion-signal
amplitude, $V$, as a function of the ionization-photon fluence,
$n$, with
\begin{equation}
V(n)=a~\frac{\sigma n}{\sigma n+ \tau_l / \tau} (1 - e^{ - \sigma
n - \tau_l/ \tau})+b, \label{Eqn:ion fun}
\end{equation}
where we set $\tau_l = 7$~ns, $\tau = 28$~ns for $5d6d~^3D_1$ and
$33$~ns for $6s7d~^3D_2$ (see APPENDIX~\ref{Section:appendix}),
$a$ is the maximum of the amplitude of the ion signal and $b$ is a
background constant (Fig.~\ref{fig:ionsignal}). The background is
from photoionization by the excitation lasers. The fluctuation of
the ion signal is due to the variation of pulse energies of the
excitation lasers leading to fluctuations in ionization
probabilities by these lasers. The number of ions detected at
highest ionization light powers is $\sim 2 \times 10^{6}$. This is
consistent with our estimate: the number of atoms in the
interaction region is $\sim 10^9$~atoms$\cdot$cm$^{-3} \times (\pi
\times 0.05^2$~cm$^2 \times 1$~cm$) \sim 10^7$~atoms, at most
one-third of them (in the case of total saturation for both
excitation transitions) are excited to the probed state, and some
of them spontaneously decay before the ionization-laser pulse
arrives.\footnote{The lifetimes of the probed states are $\sim
30$~ns (see APPENDIX~\ref{Section:appendix}) and the
ionization-laser pulse arrives the interaction region $30$~ns
later than the excitation laser pulses.} We observe that the
fluorescence signal resulting from the spontaneous decay of the
probed state drops significantly when the ionization-laser pulse
arrives in the interaction region. The polarization of the probed
state is determined by the polarizations of the excitation lasers.
The measured photoionization cross-sections with different
combinations of the polarizations of the excitation lasers are
listed in Table~\ref{table:pi cross sections}.

\begin{table}
\begin{center}
\begin{tabular}{ccccc} \hline
\hline              & \multicolumn{4}{c}{Polarizations of the excitation lasers} \\
state & $\hat{z}~\hat{z}$ & $\hat{y}~\hat{z}$ &  $\hat{z}~\hat{y}$  & $\hat{y}~\hat{y}$ \\
\hline
$5d6d~^3D_1$ & ---  & \multicolumn{2}{c}{$\sigma_0 - \sigma_2$} & --- \\
   & ---  & $ 2.2(2)$ & $1.9(2)$ & --- \\
\hline $6s7d~^3D_2$ & $\sigma_0+2\sqrt{\frac{5}{7}}\sigma_2$  &
\multicolumn{2}{c}{$\sigma_0+\sqrt{\frac{5}{7}}\sigma_2$}
& $\sigma_0-\sqrt{\frac{5}{7}}\sigma_2$ \\
  & $2.0(2)$  & $1.5(2)$ & $1.7(2)$ & $1.1(3)$\\
\hline \hline
\end{tabular}
\caption{Measured photoionization cross-sections in units of
$10^{-17}$~cm$^{2}$. The errors in the parentheses are the
statistical uncertainties. For the $5d6d~^3D_1$ state, in the
cases of both lasers polarized in the same direction, the
fluorescence signal due to spontaneous decay of the $5d6d~^3D_1$
state drops significantly and no ion signal is detected.}
\label{table:pi cross sections}
\end{center}
\end{table}

We use the polarization direction of the linearly polarized
ionization laser to define the quantization axis ($\hat{z}$). The
normalized density matrix for the ionization-laser light in the
basis of projections of the angular momentum on the quantization
axis is
\begin{equation}
\left( \begin{array}{ccc}
0 & 0 & 0 \\
0  & 1 & 0\\
0 & 0 & 0
\end{array} \right).
\label{Eqn:mixingsecular}
\end{equation}
Using Eq.~(\ref{Eqn:MMtoKQ}), we can decompose it into irreducible
tensors with components $\rho_{p~q}^{(\kappa)}$:
\begin{equation}
\begin{array}{cc}
\kappa & (q=-\kappa,\cdots,\kappa)\\
0 & \left(\frac{1}{\sqrt{3}}\right)\\
1 & \left(0,0,0\right)\\
2 & \left(0,0,-\sqrt{\frac{2}{3}},0,0\right).
\end{array}
\end{equation}

\subsection{The $5d6d~^3D_1$ state} Only the $M=\pm1$ sublevels
can be populated in our experimental setup. If one excitation
laser is polarized along the $y$-axis and the other is polarized
along the $z$-axis, the normalized density matrix in the Zeeman
basis is
\begin{equation}
\left(
\begin{array}{ccc}
\frac{1}{2}&0&-\frac{1}{2}\\
0&0&0\\
-\frac{1}{2}&0&\frac{1}{2}
\end{array}
\right).
\end{equation}
With Eq.~(\ref{Eqn:MMtoKQ}), the components of the irreducible
tensors, $\rho_{a~q}^{(\kappa)}$ are
\begin{equation}
\begin{array}{cc}
\kappa & (q=-\kappa,\cdots,\kappa)\\
0 & \left(\frac{1}{\sqrt{3}}\right)\\
1 & \left(0,0,0\right)\\
2 & \left(-\frac{1}{2},0,\frac{1}{\sqrt{6}},0,-\frac{1}{2}\right).
\end{array}
\end{equation}
From Eq.~(\ref{Eqn:cs3p}), the photoionization cross-section in
this relative polarization is
\begin{equation}
\sigma=\sigma_0 - \sigma_2.
\end{equation}
As mentioned in the end of Section~\ref{Section:appratus},
constrained by our experimental setup, this is the only
combination of photoionization cross-sections that we can
determine. As listed in Table~\ref{table:pi cross sections}, we
obtained statistically consistent photoionization cross-sections
with different polarizations of the excitation lasers. The average
cross-section is $2.0(2) \times 10^{-17}$~cm$^2$.

If we adjust the polarizations of both excitation-laser beams to
be parallel, the fluorescence signal due to the spontaneous decay
of the $5d6d~^3D_1$ state to the $6s6p~^3P_1$ state detected by
the PMT drops significantly (by more than a factor of $20$)
compared to the case of parallel polarizations and almost no ion
signal is detected. As the excitation transition is nearly
saturated when the polarization of two excitation-laser beams are
perpendicular, the residual signal may be attributed to the
imperfection of the polarizer films (polarization directions and
stray ellipticity are controlled within $5^{\circ}$.)

\subsection{The $6s7d~^3D_2$ state} Three different alignments of
this state were excited with different combinations of the
polarizations of the two excitation lasers. Following the same
approach as we used for the $5d6d~^3D_1$ state, we get that
\begin{eqnarray}
\sigma(\hat{z}\hat{z})&=&\sigma_0 +
2\sqrt{\frac{5}{7}}\sigma_2, \\
\sigma(\hat{z}\hat{y})&=&\sigma(\hat{y}\hat{z})= \sigma_0 +
\sqrt{\frac{5}{7}}\sigma_2, \\
\sigma(\hat{y}\hat{y})&=& \sigma_0 - \sqrt{\frac{5}{7}}\sigma_2.
\end{eqnarray}
Indeed, we obtained different photoionization cross-sections with
different polarizations of the excitation lasers: when both
excitation-laser beams are polarized along the $\hat{z}$-axis,
$\sigma=2.0(2)$; when both are polarized along the $\hat{y}$-axis,
$\sigma=1.1(3)$; when one along $\hat{y}$ and one along $\hat{z}$,
$\sigma=1.6(2)$ (Table~\ref{table:pi cross sections}). The fit
shows that $\sigma_0=1.3(1) \times 10^{-17}$~cm$^{2}$ and
$\sigma_2=0.43(8) \times 10^{-17}$~cm$^{2}$.
The ratio $\sigma_2/\sigma_0 \sim 0.34(7)$ suggests that this
state is coupled most to continuum states with $J=1$ or/and $3$
(Table~\ref{table:coupling J}). We are not able to derive
$\sigma_1$ because all components of the rank-one irreducible
tensor for linearly polarized ionization photons are zero. If the
ionization and excitation laser beams are circularly polarized,
$\sigma_1$ can be derived. However, this was not attempted in the
present work.


\section{Sources of Systematic error}
\label{Section:systematics}

The dominant source of the systematic error comes from our
oversimplified model of the spatial profile and the temporal
distribution of the intensity of the ionization-laser pulses. The
spatial profile of the ionization laser is approximately Gaussian,
with a diameter of $\sim 4$~mm. We measure the energy of the laser
pulse that passes through the $1.01$-mm-diameter iris and
approximate the intensity as a constant. A calculation shows that
this approximation may cause an maximum $(+ 10\%)$ correction on
the photoionization cross-section. This correction is hard to
estimate more accurately because the excitation and ionization
transitions are partially saturated. The temporal distribution of
the intensity, which is modelled as a square function, can be
actually very complicated. In our case of $\tau_l / \tau \sim 1/4$
($\tau_l \sim 7$~ns and $\tau \sim 30$~ns), a numerical
calculation shows that the variation of the fit cross-section is
within $4\%$ with several trial functions for the temporal
distribution.

A secondary source of the systematic error comes from the
measurement of the ionization-photon fluence, including the $5\%$
uncertainty on the calibration function of the photodiode used as
an energy meter and the $4\%$ uncertainty on the opening size of
the iris.

The barium sample used has natural isotopic abundance. The barium
isotopes with non-zero nuclear spin ($^{135}$Ba, $6.59\%$, and
$^{137}$Ba, $11.23\%$, both with nuclear spin $I=3/2$) have
hyperfine structure~\cite{Jit80}. We have modelled a possible
effect due to hyperfine quantum beats (see, for
example,~\cite{Haroche}) in both the $6s6p~^1P_1$ intermediate
state, and the states we photoionize. We find that this effect
from $\sim18\%$ of our Ba sample can cause a maximum $6\%$
correction on the photoionization cross-section. The correction is
hard to estimate more accurately because of the lack of the
knowledge of the temporal distribution of the ionization laser
intensity. Other sources of systematic error, including the
determination of the polarizations of the laser beams ($<2\%$) and
the nonlinearity of the photodiode ($<2\%$), are found to be
negligible. Overall, all the photoionization cross-sections
measured in this work have a systematic uncertainty of $\sim
13\%$. In Table~\ref{table:final results}, we have listed all
experimental results measured in this work with systematic and
statistical errors.

\begin{table}
\begin{center}
\begin{tabular}{cl}

\hline \hline

state & photoionization cross-sections \\
\hline
$5d6d~^3D_1$ & $\sigma_0 - \sigma_2 = 2.0(3)\times 10^{-17}$~cm$^2$\\
\hline
$6s7d~^3D_2$ & $\sigma_0=1.3(2)\times 10^{-17}$~cm$^2$\\
& $\sigma_2=0.43(9)\times 10^{-17}$~cm$^2$\\
& $\sigma_2/\sigma_0=0.34(7)$\\
\hline \hline
\end{tabular}
\caption{Measured photoionization cross-sections with systematic
and statistical errors combined.} \label{table:final results}
\end{center}
\end{table}

\section{Conclusion}

In this work, we have measured the photoionization cross-sections
of the $5d6d~^3D_1$ and $6s7d~^3D_2$ states of Ba with the
ionization-laser wavelength $556.6$~nm. We found that the
photoionization cross-section of the $6s7d~^3D_2$ state depends on
the relative polarizations of the atomic state and the
ionization-laser beam. We have introduced a general tensor
formalism of polarization-dependent photoionization cross-sections
and determined two of the three parameters of the photoionization
cross-section of the $6s7d~^3D_2$ state.

\bigskip

\acknowledgments

The authors wish to thank D. English and S. M. Rochester for help
with the experiments and useful discussions, and D. Angom, M.
Auzinsh, R. deCarvalho, M. Havey, J. Higbie, D. Kleppner, M. G.
Kozlov and J. E. Stalnaker for helpful advice. This research was
supported by the National Science Foundation.

\appendix

\section{Radiative lifetime measurement}
\label{Section:appendix}

In section~\ref{Section:results}, it has been shown that the
temporal evolution of ion signals studied in this work depends on
the radiative lifetimes of the probed states. Using almost the
same experimental setup, we have measured the radiative lifetimes
of five even-parity excited states of Ba.

The barium atoms in an atomic beam, with estimated density of
$\sim 10^{9}$~atoms$/$cm$^3$ in the interaction region, are
excited by pulsed lasers to the even-parity states of interest via
two E1 transitions. For different probed states, different
combinations of the lasers, including a frequency-doubled Nd:YAG
laser (wavelength $\sim 532$~nm) and dye lasers with Rhodamine~6G
(wavelength $556-570$~nm) or Fluorescein~548 (wavelength
$546-567$~nm), are used for the excitation. Some states of
interest may be probed by different combinations of lasers as a
check of consistency. In some cases of lifetime measurement, atoms
can be first excited to the $6s6p~^1P_1$ state efficiently by the
amplified spontaneous emission (ASE) of the dye laser because the
transition probability of $6s^2~^1S_0 \rightarrow 6s6p~^1P_1$ is
large; therefore, only one dye laser is used to excite a two-step
E1-E1 transition.

Fluorescence resulting from spontaneous decay to a lower-lying
odd-parity state was detected with a PMT. A colored-glass filter
was used to reduce scattered light from the lasers, interference
filters with $10$-nm bandwidth were used to select the decay
channel of interest and a linear-polarizing film was used to
select a polarization of the fluorescence. We recorded the
time-dependent fluorescence signals with a digital oscilloscope
and analyzed data with a personal computer running the Mathematica
program. We recorded fluorescence signals without averaging
because we found that the averaging in general elongates the
apparent lifetime and the fitted lifetime is sensitive to the
number of the averaged samples. This is probably the result of
jitter in the triggering of the oscilloscope and/or the lasers.

\begin{figure}
\includegraphics[width=3.25 in]{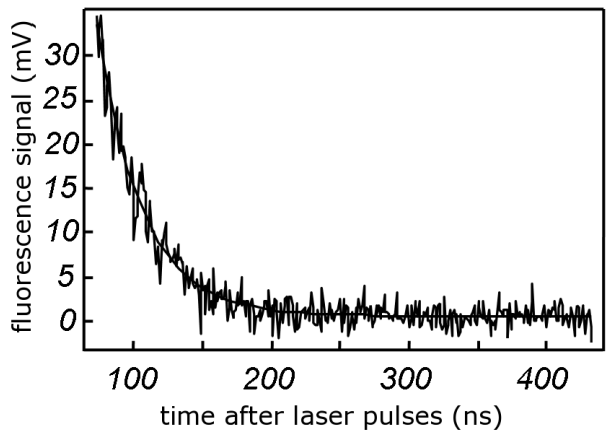}
\caption{Fluorescence signal due to spontaneous decay of the
$6s7d~^3D_2$ state. The data points with time later than $80$~ns
are fit to an exponential function.} \label{fig:decayplot}
\end{figure}

We determine the radiative lifetime of a probed state by fitting
the fluorescence signal due to spontaneous decay with an
exponential function:
\begin{equation}
f(t)= a~e^{-t/\tau}+b, {\rm ~for~} t \ge 0,
\end{equation}
where $a$ is the signal amplitude, $\tau$ is the radiative
lifetime of the probed state, $b$ is the constant background and
the probed state is populated at $t=0$ (Fig.~\ref{fig:decayplot}).
It can be shown (APPENDIX~\ref{pmtresponse}) that we can avoid the
effects of the finite PMT response time, the finite laser pulse
width and the finite oscilloscope response time if only the data
points with time sufficiently long after the laser excitation are
used in the fitting.

\begin{figure}
\includegraphics[width=3.25 in]{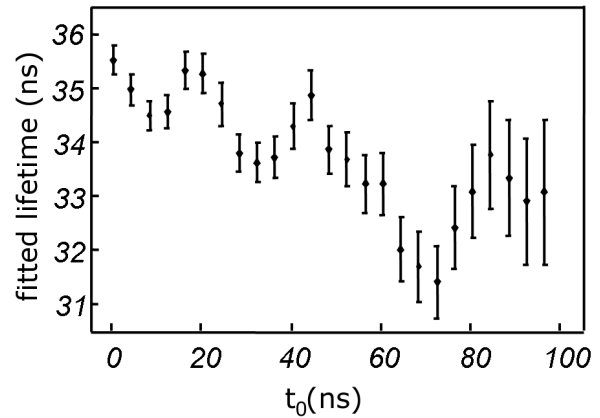}
\caption[Fitted lifetime as a function of $t_0$]{A typical plot of
fitted lifetime as a function of $t_0$. The error bar is one
standard deviation of statistical uncertainty. The fitted lifetime
varies as $t_0$ increases. Typically, when $t_0 > 75$~ns, the
fitting gives a consistent lifetime, which means that the effect
of PMT response, finite temporal width of laser pulse etc. is
negligible.} \label{fig:lifetime}
\end{figure}

In the data analysis, only data points later than a certain time,
$t_0 (> 0)$, were fit to an exponential function. The fitted
lifetime may vary with $t_0$ if $t_0$ is not sufficiently long. As
$t_0$ increases, the fitted lifetime will approach a consistent
value, which means that the effects of PMT response etc. become
negligible (Fig.~\ref{fig:lifetime}). If $t_0$ is chosen too long,
there is no signal left to fit. Typically, we found that when
$t_0$ is greater than $\sim 75$~ns, the fitting gives a consistent
lifetime.

The lifetimes determined in this work are listed in
Table~\ref{table:lifetime}. The lifetimes of the same probed
states were measured from the fluorescence signals with different
excitation schemes (different combinations of lasers or ASE) and
different detection schemes (different IFs and/or different
orientation of the polarizing film). The fitting gives consistent
results. Comparing this work with previous experiments, we found
that the lifetimes of the $6s7d~^3D_2$ and $6s7d~^3D_1$ states
disagree with those reported in Ref.~\cite{Sme93} by more than two
standard deviations. The lifetimes of $5d6d~^3S_1$ and
$5d6d~^3D_1$ states agree with the previous results within one
standard deviation~\cite{Sme93, DerUndthesis}.

\begin{table}
\begin{center}
\begin{tabular}{cccc} \hline
\hline &    & \multicolumn{2}{c}{Lifetime (ns)} \\
Upper state & Lower state &  this work & previous work  \\
\hline
$6s7d~^3D_2$  & $6s6p~^3P_1$   &  33(2) & 39.0(18)~\cite{Sme93}  \\

$6s7d~^3D_1$  & $6s6p~^3P_1$  &   34(2)  & 39.0(12)~\cite{Sme93}   \\  

$5d6d~^3S_1$  & $6s6p~^3P_2$    &  25(2) & 25(15)~\cite{DerUndthesis} \\  

$5d6d~^3D_1$  & $6s6p~^3P_1$ &   28(5)   &  23.0(18)~\cite{Sme93}  \\

$5d6d~^3D_2$  & $6s6p~^3P_2$  &  23(2)   &  \\  
\hline \hline
\end{tabular}
\caption{Radiative lifetimes determined in this work. The
presented errors of lifetimes determined in this work include
statistical and systematic uncertainties.} \label{table:lifetime}
\end{center}
\end{table}

Hyperfine quantum beats are a potential source of systematic error
for lifetime measurements. Our simulation shows that there can be
a maximum $3\%$ systematic error on the radiative lifetime if we
fit the fluorescence signal with an exponential function. Other
sources of systematic errors may result from the finite response
time, afterpulses and nonlinearity of the PMT. We found that these
systematic uncertainties can be minimized $< 1\%$, which is much
smaller than statistical uncertainties, by appropriate
experimental procedure. To avoid any possible detection of
unexpected cascade fluorescence channels with wavelength
``coincidentally" close to the target fluorescence, we have
searched all the possible decay transitions according to the
latest updated energy levels of neutral
barium~\cite{Cur2004}.\footnote{All the energy levels of barium
below the probe states have been identified except the
$5d^2~^1G_4$ state. Transitions to this state from the levels of
interest are forbidden by the angular-momentum selection rules.}
No cascade decay channels of the probed state were found to be
detectable. We have also used different IFs for the same decay
channels. They all give a consistent lifetime. In the data
analysis, we also subtract the temporal fluorescence data by
off-resonance data to eliminate the effect of the scattered light
and any off-resonance interactions.

\section{PMT and Oscilloscope Response} \label{pmtresponse}

The PMT used in this work has a rise time of $\sim 8$~ns and a
response time of $\sim 15$~ns FWHM. The oscilloscope bandwidth is
$500$~MHz. The temporal width of the laser pulse is about $7$~ns.
The lifetimes of the probed states are all less than $40$~ns.
Therefore, the systematic effect due to these finite responses
should be considered. We prove that this effect becomes negligible
if only data points with time sufficiently long after the laser
pulses are considered.

The fluorescence signal due to a spontaneous decay can be
expressed as an exponential function:
\begin{equation}
F(t)= \left\{ \begin{array}{ll} 0, & {\rm ~for~} t<0\\
e^{-t/\tau},& {\rm ~for~} t\ge0
\end{array}, \right.
\label{Eqn:decay}
\end{equation}
where $\tau$ is the radiative lifetime of the probed state and the
probed state is prepared at $t=0$.

Assume that after a sharp light pulse, the PMT signal follows an
exponential decay function. We can model the PMT response with the
following function:
\begin{equation}
R(t)= \left\{ \begin{array}{ll} 0, & {\rm ~for~} t<0\\
P(t)\ e^{-t/\tau_p},& {\rm ~for~} t>0
\end{array}, \right.
\label{Eqn:PMT response}
\end{equation}
where $P$ is any polynomial and $\tau_p$ is the characteristic
time of the PMT response. In this work, we have the PMT response
that decays faster than the fluorescence, i.e., $\tau_p < \tau$.

The signal observed on the oscilloscope, $S(t)$, is the
convolution of fluorescence signal with the PMT response function:
\begin{eqnarray}
S(t)&=& \int_0^t F(t')R(t-t')dt' \nonumber \\
&=& \int_0^t e^{-t'/\tau}~R(t-t')dt' \nonumber\\
&=& \int_{-\infty}^t e^{-t'/\tau}~R(t-t')dt' \nonumber\\
&&-\int_{-\infty}^0 e^{-t'/\tau}~R(t-t')dt' \nonumber\\
& \equiv & S_1(t)-S_2(t). \label{Eqn:PMT+decay}
\end{eqnarray}
To simplify $S_1(t)$, we change the integrated variable $t'$ to
$t-x$:
\begin{eqnarray}
S_1(t)& = & \int_0^\infty e^{(x-t)/\tau}~R(x)dx \nonumber\\
      & = &
      e^{-t/\tau}\int_0^{\infty}e^{x(1/\tau-1/\tau_p)}P(x)dx.
\label{Eqn:S1-1}
\end{eqnarray}
Because $\tau_p$ is smaller than $\tau$, the integral in
Eq.(\ref{Eqn:S1-1}) converges and is a finite constant. Therefore,
\begin{equation}
S_1(t)=C~e^{-t/\tau}. \label{Eqn:S1-2}
\end{equation}
The function $S_2(t)$ can also be simplified,
\begin{eqnarray}
S_2(t)&=&\int_{-\infty}^0
e^{-t'/\tau}~P(t-t')~e^{-(t-t')/\tau_p}dt' \nonumber \\
&=& e^{-t/\tau_p} \int_{-\infty}^0
e^{-t'(1/\tau-1/\tau_p)}~P(t-t')dt'. \label{Eqn:S2-1}
\end{eqnarray}
Because $\tau_p$ is smaller than $\tau$, the integral in
Eq.(\ref{Eqn:S2-1}) converges and is another polynomial. Hence, we
get
\begin{equation}
S_2(t)=P'(t)~e^{-t/\tau_p}. \label{Eqn:S2-2}
\end{equation}
Using Eq.~(\ref{Eqn:S1-2}) and Eq.~(\ref{Eqn:S2-2}), we can
simplify the observed signal function as
\begin{equation}
S(t)=C~e^{-t/\tau}-P'(t)~e^{-t/\tau_p}. \label{Eqn:S}
\end{equation}
As $t$ becomes large, the second term in Eq.~(\ref{Eqn:S})
approaches zero faster than the first term; therefore, the effect
of the PMT response becomes negligible.

\bigskip

\bibliography{Ba_lifetime_cs}

\begin{thebibliography}{16}
\expandafter\ifx\csname natexlab\endcsname\relax\def\natexlab#1{#1}\fi
\expandafter\ifx\csname bibnamefont\endcsname\relax
  \def\bibnamefont#1{#1}\fi
\expandafter\ifx\csname bibfnamefont\endcsname\relax
  \def\bibfnamefont#1{#1}\fi
\expandafter\ifx\csname citenamefont\endcsname\relax
  \def\citenamefont#1{#1}\fi
\expandafter\ifx\csname url\endcsname\relax
  \def\url#1{\texttt{#1}}\fi
\expandafter\ifx\csname urlprefix\endcsname\relax\def\urlprefix{URL }\fi
\providecommand{\bibinfo}[2]{#2}
\providecommand{\eprint}[2][]{\url{#2}}

\bibitem[{\citenamefont{Kelly}(1990)}]{Kel90}
\bibinfo{author}{\bibfnamefont{H.~P.} \bibnamefont{Kelly}}, in
  \emph{\bibinfo{booktitle}{AIP Conference Proceedings}}
  (\bibinfo{year}{1990}), vol. \bibinfo{volume}{215} of
  \emph{\bibinfo{series}{AIP Conf. Procs.}}, p. \bibinfo{pages}{292}.

\bibitem[{\citenamefont{Lubell and Raith}(1969)}]{Lub69}
\bibinfo{author}{\bibfnamefont{M.~S.} \bibnamefont{Lubell}} \bibnamefont{and}
  \bibinfo{author}{\bibfnamefont{W.}~\bibnamefont{Raith}},
  \bibinfo{journal}{Phys. Rev. Lett.} \textbf{\bibinfo{volume}{23}},
  \bibinfo{pages}{211} (\bibinfo{year}{1969}).

\bibitem[{\citenamefont{Fox et~al.}(1971)\citenamefont{Fox, Kogan, and
  Robinson}}]{Fox71}
\bibinfo{author}{\bibfnamefont{R.~A.} \bibnamefont{Fox}},
  \bibinfo{author}{\bibfnamefont{R.~M.} \bibnamefont{Kogan}}, \bibnamefont{and}
  \bibinfo{author}{\bibfnamefont{E.~J.} \bibnamefont{Robinson}},
  \bibinfo{journal}{Phys. Rev. Lett.} \textbf{\bibinfo{volume}{26}},
  \bibinfo{pages}{1416} (\bibinfo{year}{1971}).

\bibitem[{\citenamefont{Kogan et~al.}(1971)\citenamefont{Kogan, Fox, Burnham,
  and Robinson}}]{Kog71}
\bibinfo{author}{\bibfnamefont{R.~M.} \bibnamefont{Kogan}},
  \bibinfo{author}{\bibfnamefont{R.~A.} \bibnamefont{Fox}},
  \bibinfo{author}{\bibfnamefont{G.~T.} \bibnamefont{Burnham}},
  \bibnamefont{and} \bibinfo{author}{\bibfnamefont{E.~J.}
  \bibnamefont{Robinson}}, \bibinfo{journal}{Bull. Amer. Phys. Soc.}
  \textbf{\bibinfo{volume}{16}}, \bibinfo{pages}{1411} (\bibinfo{year}{1971}).

\bibitem[{\citenamefont{DeMille et~al.}(1999)\citenamefont{DeMille, Budker,
  Derr, and Deveney}}]{Dem99}
\bibinfo{author}{\bibfnamefont{D.}~\bibnamefont{DeMille}},
  \bibinfo{author}{\bibfnamefont{D.}~\bibnamefont{Budker}},
  \bibinfo{author}{\bibfnamefont{N.}~\bibnamefont{Derr}}, \bibnamefont{and}
  \bibinfo{author}{\bibfnamefont{E.}~\bibnamefont{Deveney}},
  \bibinfo{journal}{Phys. Rev. Lett.} \textbf{\bibinfo{volume}{83}},
  \bibinfo{pages}{3978} (\bibinfo{year}{1999}).

\bibitem[{\citenamefont{English et~al.}(2000)\citenamefont{English, Budker, and
  DeMille}}]{Eng00}
\bibinfo{author}{\bibfnamefont{D.}~\bibnamefont{English}},
  \bibinfo{author}{\bibfnamefont{D.}~\bibnamefont{Budker}}, \bibnamefont{and}
  \bibinfo{author}{\bibfnamefont{D.}~\bibnamefont{DeMille}}, in
  \emph{\bibinfo{booktitle}{proceedings of the International Conference on
  Spin-Statistics Connection and Commutation Relations: Experimental Tests and
  Theoretical Implications}}, edited by \bibinfo{editor}{\bibfnamefont{R.~C.}
  \bibnamefont{Hilborn}} \bibnamefont{and}
  \bibinfo{editor}{\bibfnamefont{G.~M.} \bibnamefont{Tino}}
  (\bibinfo{address}{Anacapri, Italy}, \bibinfo{year}{2000}), vol.
  \bibinfo{volume}{545} of \emph{\bibinfo{series}{AIP Conf. Procs.}}, p.
  \bibinfo{pages}{281}.

\bibitem[{\citenamefont{Rochester et~al.}(1999)\citenamefont{Rochester, Bowers,
  Budker, DeMille, and Zolotorev}}]{Roc99}
\bibinfo{author}{\bibfnamefont{S.~M.} \bibnamefont{Rochester}},
  \bibinfo{author}{\bibfnamefont{C.~J.} \bibnamefont{Bowers}},
  \bibinfo{author}{\bibfnamefont{D.}~\bibnamefont{Budker}},
  \bibinfo{author}{\bibfnamefont{D.}~\bibnamefont{DeMille}}, \bibnamefont{and}
  \bibinfo{author}{\bibfnamefont{M.}~\bibnamefont{Zolotorev}},
  \bibinfo{journal}{Phys. Rev. A.} \textbf{\bibinfo{volume}{59}},
  \bibinfo{pages}{3480} (\bibinfo{year}{1999}).

\bibitem[{\citenamefont{Li et~al.}(2004)\citenamefont{Li, Rochester, Kozlov,
  and Budker}}]{Li04}
\bibinfo{author}{\bibfnamefont{C.-H.} \bibnamefont{Li}},
  \bibinfo{author}{\bibfnamefont{S.~M.} \bibnamefont{Rochester}},
  \bibinfo{author}{\bibfnamefont{M.~G.} \bibnamefont{Kozlov}},
  \bibnamefont{and} \bibinfo{author}{\bibfnamefont{D.}~\bibnamefont{Budker}},
  \bibinfo{journal}{Phys. Rev. A.} \textbf{\bibinfo{volume}{69}},
  \bibinfo{pages}{042507} (\bibinfo{year}{2004}).

\bibitem[{\citenamefont{Alexandrov et~al.}(2005)\citenamefont{Alexandrov,
  Auzinsh, Budker, Kimball, Rochester, and Yashchuk}}]{Ale2005}
\bibinfo{author}{\bibfnamefont{E.~B.} \bibnamefont{Alexandrov}},
  \bibinfo{author}{\bibfnamefont{M.}~\bibnamefont{Auzinsh}},
  \bibinfo{author}{\bibfnamefont{D.}~\bibnamefont{Budker}},
  \bibinfo{author}{\bibfnamefont{D.~F.} \bibnamefont{Kimball}},
  \bibinfo{author}{\bibfnamefont{S.~M.} \bibnamefont{Rochester}},
  \bibnamefont{and} \bibinfo{author}{\bibfnamefont{V.~V.}
  \bibnamefont{Yashchuk}}, \bibinfo{journal}{JOSA B}
  \textbf{\bibinfo{volume}{22(1)}}, \bibinfo{pages}{7} (\bibinfo{year}{2005}).

\bibitem[{\citenamefont{Blum}(1996)}]{Blum}
\bibinfo{author}{\bibfnamefont{K.}~\bibnamefont{Blum}},
  \emph{\bibinfo{title}{Density Matrix Theory and Applications}}
  (\bibinfo{publisher}{Plenum Publishing Corporation}, \bibinfo{address}{233
  Spring Street, New York, NY 10013}, \bibinfo{year}{1996}).

\bibitem[{\citenamefont{Sobel'man}(1992)}]{Sobelman}
\bibinfo{author}{\bibfnamefont{I.~I.} \bibnamefont{Sobel'man}},
  \emph{\bibinfo{title}{Atomic spectra and radiative transitions}}
  (\bibinfo{publisher}{Springer-Verlag}, \bibinfo{address}{Berlin},
  \bibinfo{year}{1992}).

\bibitem[{\citenamefont{Jitschin and Meisel}(1980)}]{Jit80}
\bibinfo{author}{\bibfnamefont{W.}~\bibnamefont{Jitschin}} \bibnamefont{and}
  \bibinfo{author}{\bibfnamefont{G.}~\bibnamefont{Meisel}},
  \bibinfo{journal}{Zeitschrift fur Physik A} \textbf{\bibinfo{volume}{295}},
  \bibinfo{pages}{37} (\bibinfo{year}{1980}).

\bibitem[{\citenamefont{Haroche}(1976)}]{Haroche}
\bibinfo{author}{\bibfnamefont{S.}~\bibnamefont{Haroche}}, in
  \emph{\bibinfo{booktitle}{High-resolution Laser Spectroscopy}}, edited by
  \bibinfo{editor}{\bibfnamefont{K.}~\bibnamefont{Shimoda}}
  (\bibinfo{publisher}{Springer-Verlag}, \bibinfo{address}{Berlin.},
  \bibinfo{year}{1976}), p. \bibinfo{pages}{256}.

\bibitem[{\citenamefont{Smedley and Marran}(1993)}]{Sme93}
\bibinfo{author}{\bibfnamefont{J.~E.} \bibnamefont{Smedley}} \bibnamefont{and}
  \bibinfo{author}{\bibfnamefont{D.~F.} \bibnamefont{Marran}},
  \bibinfo{journal}{Phys. Rev. A} \textbf{\bibinfo{volume}{47}},
  \bibinfo{pages}{126} (\bibinfo{year}{1993}).

\bibitem[{\citenamefont{Derr}(1997)}]{DerUndthesis}
\bibinfo{author}{\bibfnamefont{N.}~\bibnamefont{Derr}},
  \bibinfo{journal}{Undergraduate thesis, UC Berkeley.}
  (\bibinfo{year}{1997}).

\bibitem[{\citenamefont{Curry}(2004)}]{Cur2004}
\bibinfo{author}{\bibfnamefont{J.~J.} \bibnamefont{Curry}},
  \bibinfo{journal}{J. Phys. Chem. Ref. Data} \textbf{\bibinfo{volume}{33}},
  \bibinfo{pages}{725} (\bibinfo{year}{2004}).

\end{thebibliography}

\end{document}